# Confluent Hypergeometric Equation via Fractional Calculus Approach

Rodrigues FG* and Capelas de Oliveira E

*Department of Applied Mathematics, IMECC – UNICAMP, Brazil*

**Abstract**

In this paper, using the theory of the so-called fractional calculus we show that it is possible to easily obtain the solutions for the confluent hypergeometric equation. Our approach is to be compared with the standard one (Frobenius) which is based on the ordinary calculus of integer order.

## Introduction

Investigations in the so-called fractional calculus, that is, the theory of integration and differentiation of arbitrary order, have increased considerably in the last fifty years [1-7]. Nowadays, the topic is broadly dispersed in several distinct applications in science such as fractional viscoelasticity [8,9], fractional harmonic oscillators [9,10], fractional frictional forces [10], fractional fluid dynamics [11] and fractional signal processing [12,13], just to mention a few examples. Such investigations indicates that the descriptions of previously established theories by means of ordinary calculus (of integer order) are now being reviewed by means of fractional calculus with promising results such as the increase in accuracy between theoretical and experimental data.

Following this trend of approaching "old problems" by means of "new tools", recently we have approached Bessel's differential equation of order p [14] by a methodology of the fractional calculus as proposed by the authors [15]. Although we were successful in obtaining Bessel's solutions of the first kind, the methodology itself was not rigorously accurate (from a mathematical point of view) to justify the second linear independent solution, specifically $J_p$ for $0 < p \neq -1, -2,...$ inspiring us to continue with other investigations about the methodology by applying it to a more general setting, the confluent hypergeometric equation. Also, as conjectured in the final considerations [14], we have pointed out that there might be some "minor patches" needed to be done to the definitions of the fractionalized versions of the integral and differential operators we have used in that paper (which was the Riemann- Liouville formulation and that we will present it in the next section of this paper). Nevertheless, the fractional methodology is still very promising and therefore we continue its investigations and, as mentioned above, we now apply it to obtain solutions of the confluent hypergeometric equation.

The paper is organized as follows: In section 2 we review the Riemann-Liouville integrodifferential operator particularly presenting an alternative definition by means of the Hadamard's finite part integral. In section 3 we discuss the confluent hypergeometric equation be means of the fractional methodology. In section 4 we present our conclusions and some remarks.

## Riemann-Liouville Integrodifferential Operator

In this section we describe the Riemann-Liouville formulations; hence we will present two main definitions of the fractional integral operator and fractional differential operator [16-20]. Some properties of those operators are also discussed.

**Definition 1:** Let $\Omega = [a, b] \subset \mathbb{R}$ be a finite real interval. Suppose $f \in L_1[a, b]$ is a Lebesgue integrable function in $[a, b]$. Then the expressions $\mathcal{I}_{a+}^{\nu} f$ and $\mathcal{I}_{b-}^{\nu} f$ established by the equalities below

$$\left(\mathcal{I}_{a+}^{\nu} f\right)(x) \equiv \frac{1}{\Gamma(\nu)} \int_a^x (x-t)^{\nu-1} f(t)\,dt, \qquad (1)$$

with $x > a$, $\operatorname{Re}(\nu) > 0$ and

$$\left(\mathcal{I}_{b-}^{\nu} f\right)(x) \equiv \frac{1}{\Gamma(\nu)} \int_x^b (t-x)^{\nu-1} f(t)\,dt, \qquad (2)$$

with $x < b$, $\operatorname{Re}(\nu) > 0$ and where $\Gamma(\nu)$ is the gamma function, defines the Riemann-Liouville fractional integral operator (RLFI) of order $\nu \in \mathbb{C}$. The integrals in Equation (1) and Equation (2) are usually called the left and right versions, respectively.

**Definition 2:** Let $\Omega = [a, b] \subset \mathbb{R}$ be a finite real interval. Suppose $f \in AC^n[a, b]$ is an absolute continuous function until order $\mathbf{n} - 1$. Then the eepressions $\mathcal{I}_{a+}^{\nu} f$ and $\mathcal{I}_{b-}^{\nu} f$ established by the equalities below

$$\left(\mathcal{D}_{a+}^{\nu} f\right)(x) \equiv \mathcal{D}_{a+}^{\mathbf{n}}\left[\left(\mathcal{I}_{a+}^{\mathbf{n}-\nu} f\right)(x)\right] = \left(\frac{d}{dx}\right)^{\mathbf{n}} \left(\mathcal{I}_{a+}^{\mathbf{n}-\nu} f\right)(x)$$

$$= \frac{1}{\Gamma(\mathbf{n}-\nu)} \left(\frac{d}{dx}\right)^{\mathbf{n}} \int_a^x (x-t)^{\mathbf{n}-\nu-1} f(t)\,dt, \qquad (3)$$

with $x > a$ and

$$\left(\mathcal{D}_{b-}^{\nu} f\right)(x) \equiv \mathcal{D}_{a+}^{\mathbf{n}}\left[\left(\mathcal{I}_{b-}^{\mathbf{n}-\nu} f\right)(x)\right] = \left(-\frac{d}{dx}\right)^{\mathbf{n}} \left(\mathcal{I}_{b-}^{\mathbf{n}-\nu} f\right)(x)$$

$$= \frac{1}{\Gamma(\mathbf{n}-\nu)} \left(-\frac{d}{dx}\right)^{\mathbf{n}} \int_x^b (t-x)^{\mathbf{n}-\nu-1} f(t)\,dt, \qquad (4)$$

with $x < b$ and $[\operatorname{Re}(\nu)]$ being the integer part function of $\operatorname{Re}(\nu)$ and $\mathbf{n} = [\operatorname{Re}(\nu)] + 1$, defines the Riemann-Liouville fractional differential operators (RLFD) of order $\nu \in \mathbb{C}$ $(\operatorname{Re}(\nu) \geq 0)$. The derivatives in Equation (1) and Equation (2) are usually called the left and right versions, respectively.

---

Before continuing with the presentation of new definitions and results, we inform the reader that in this work we will restrict ourselves to the left versions (Equation (1) and Equation (3)) only, and consider the order v of the operators to be a real number.

Mind that the two definitions are reduced to the classical integer cases whenever we choose $v = n \in \mathbb{N}$. In particular, it can be shown [16,20] that if $v = n = 0$, then we have the identity operator **I**:

$$\mathcal{D}_{a+}^{0} = \mathbf{I} = \mathcal{I}_{a+}^{0} := \lim_{v \to 0} \mathcal{I}_{a+}^{v}.$$

We also point out that if $f(x)$ is of the form $(x - a)^{\beta - 1}$ for $\beta > 0$, then [17]

$$\left(\mathcal{I}_{a+}^{v}(t-a)^{\beta-1}\right)(x) = \frac{\Gamma(\beta)}{\Gamma(\beta+v)}(x-a)^{\beta+v-1}, \quad (5a)$$

$$\left(\mathcal{D}_{a+}^{v}(t-a)^{\beta-1}\right)(x) = \frac{\Gamma(\beta)}{\Gamma(\beta-v)}(x-a)^{\beta-v-1}. \quad (5b)$$

As a consequence of these rules and the properties of $\mathcal{I}_{a+}^{v}$ and $\mathcal{D}_{a+}^{v}$ are continuous operators relative to the index $v$ [16], then if $f(x) \in C^{\infty}[a, \epsilon]$ is analytic in the interval $[a, \epsilon]$ for some $\epsilon > a$, then $f$ is representable as a power series

$$f(x) = \sum_{k=0}^{\infty} c_k (x-a)^k, \quad (6)$$

where $c_k$ are constants and, in this case, applying the rules as in Equation (5a) and Equation (5b) we may integro differentiate functions as per Equation (6) according to the formulas [15,16,20]:

$$\left[\mathcal{I}_{a+}^{v} \sum_{k=0}^{\infty} c_k (t-a)^k\right](x) = \sum_{k=0}^{\infty} c_k \mathcal{I}_{a+}^{v}(x-a)^k = \sum_{k=0}^{\infty} c_k \frac{\Gamma(k+1)(x-a)^{k+v}}{\Gamma(k+1+v)} \quad (7)$$

and

$$\left[\mathcal{D}_{a+}^{v} \sum_{k=0}^{\infty} c_k (t-a)^k\right](x) = \sum_{k=0}^{\infty} c_k \mathcal{D}_{a+}^{v}(x-a)^k = \sum_{k=0}^{\infty} c_k \frac{\Gamma(k+1)(x-a)^{k-v}}{\Gamma(k+1-v)}. \quad (8)$$

We now present a "unified" version of the RLFI and RLFD under a single operator symbol $\mathfrak{D}_{a+}^{v}$. We shall call it the Riemann-Liouville fractional integro differential operator or simply integro differential operator. That is, for $v \in \mathbb{R}$ and assuming that $f$ is such that $\mathcal{D}_{a+}^{v} f$ and $\mathcal{I}_{a+}^{v} f$ are well defined, then

$$\mathfrak{D}_{a+}^{v} = \begin{cases} \mathcal{D}_{a+}^{v}, & v > 0 \text{ (RLFD)}, \\ \mathbf{I}, & v = 0, \\ \mathcal{I}_{a+}^{-v}, & v < 0 \text{ (RLFI)}, \end{cases} \quad (9)$$

so that we have the following situation for $v > 0$

$$\left(\mathcal{D}_{a+}^{-v} f\right)(x) = \left(\mathcal{I}_{a+}^{v} f\right)(x), \quad (10)$$

and

$$\left(\mathcal{D}_{a+}^{v} f\right)(x) = \left(\mathcal{I}_{a+}^{-v} f\right)(x). \quad (11)$$

This suggests that the RLFD can be interpreted as an analytical extension of the operator RLFI. However we must take some caution here. As we have pointed out [14], the integral

$$\left(\mathcal{I}_{a+}^{-v} f\right)(x) = \frac{1}{\Gamma(-v)} \int_{a}^{x} (x-t)^{-v-1} f(t) \mathbf{d}t,$$

usually diverges even if $f(x) \in L_1[a, b]$. To avoid this "pitfall" we need to consider the so-called Hadamard's finite part integral [16,21]:

$$\mathcal{H} \int_{a}^{b} (x-a)^{-\beta} f(x) \mathbf{d}x := \sum_{k=0}^{\mathbf{n}} \frac{f^{(k)}(a)(b-a)^{k+1-\beta}}{(k+1-\beta)k!} + \int_{a}^{b} (x-a)^{-\beta} R_{\mathbf{n}}(x,a) \mathbf{d}x, \quad (12)$$

where $\mathbf{n} = [\beta] + 1$, $\beta \notin \mathbb{N}$ and

$$R_{\mathbf{n}}(x,a) = \frac{1}{\mathbf{n}!} \int_{a}^{x} (x-t)^{\mathbf{n}} f^{(\mathbf{n}+1)}(t) \mathbf{d}t$$

is the remainder of the nth degree Taylor polynomial of $f$ expanded at $x = a$.

An alternative definition for the Hadamard's finite part integral can be stated in terms of the following theorem [16].

**Theorem 3:** Let $1 < \beta \notin \mathbb{N}$ and $\mathbf{m} = [\beta] + 1$. For $f \in C^m[a, b]$ is $\mathbf{m}$ times continuously differentiable, then

$$\frac{1}{\Gamma(1-\beta)} \mathcal{H} \int_{a}^{b} (x-a)^{-\beta} f(x) \mathbf{d}x = \sum_{k=0}^{\mathbf{m}-1} \frac{f^{(k)}(a)(b-a)^{k+1-\beta}}{\Gamma(k+2-\beta)} + \mathcal{I}_{a+}^{\mathbf{m}-\beta+1} f^{(\mathbf{m})}(b).$$

In view of this result, we can interpret the RLFD as an integral as stated in the next theorem [16].

**Theorem 4:** Let $0 < v \notin \mathbb{N}$ and $n = [v] + 1$. If $f \in C^n[a, b]$ and $x \in [a, b]$, then

$$\left(\mathcal{D}_{a+}^{v} f\right)(x) = \frac{1}{\Gamma(-v)} \mathcal{H} \int_{a}^{x} (x-t)^{-v-1} f(t) \mathbf{d}t$$

So in fact, the formal interpretation of Equation (10) and Equation (11) must be

$$\left(\mathcal{D}_{a+}^{v} f\right)(x) = \mathcal{H}\left(\mathcal{I}_{a+}^{-v} f\right)(x), \quad (13)$$

$$\left(\mathcal{D}_{a+}^{-v} f\right)(x) = \mathcal{H}\left(\mathcal{I}_{a+}^{v} f\right)(x), \quad (14)$$

for $v \neq 1, 2, \ldots$. Nevertheless, for simplicity in notation, we shall use the ones established in Equation (10) and Equation (11), but having the interpretations as in Equation (13) and Equation (14).

We now summarize, without presenting the proofs[1], some composition laws for the $\mathfrak{D}_{a+}^{v}$ operator that will be needed for the next section.

The first of them is the following:

$$\mathfrak{D}_{a+}^{-p}\left[\mathfrak{D}_{a+}^{q} f(x)\right] = \mathfrak{D}_{a+}^{q-p} f(x) - \sum_{k=0}^{\mathbf{n}-1} \frac{(\mathfrak{D}_{a+}^{q-k-1} f)(a)}{\Gamma(p-k)}(x-a)^{p-k-1}, \quad (15)$$

where $p, q \in \mathbb{R}^+$ and $\mathbf{n} = [q] + 1$.

Then for $v > 0$ and $m \in \mathbb{N}$, we have

$$\left(\mathfrak{D}_{a+}^{m} \mathfrak{D}_{a+}^{v} f\right)(x) = \left(\mathfrak{D}_{a+}^{m+v} f\right)(x), \quad (16)$$

$$\left(\mathfrak{D}_{a+}^{v} \mathfrak{D}_{a+}^{m} f\right)(x) = \left(\mathfrak{D}_{a+}^{m+v} f\right)(x) - \sum_{k=0}^{m-1} \frac{f^{(k)}(a)(x-a)^{k-v-m}}{\Gamma(1+k-v-m)}. \quad (17)$$

Also, assuming that the derivatives exists, we have the equalities

$$\mathfrak{D}_{a+}^{p}\left[\mathfrak{D}_{a+}^{q} f(x)\right] = \mathfrak{D}_{a+}^{p+q} f(x) - \sum_{k=0}^{\mathbf{m}-1} \frac{(\mathfrak{D}_{a+}^{q-k-1} f)(a)}{\Gamma(-p-k)}(x-a)^{-p-k-1}, \quad (18)$$

---

[1]The proofs can be found in several distinct sources. We mention the main literature where these results can be found, [16,17,19,20].

$$\mathfrak{D}_{a+}^{q}\left[\mathfrak{D}_{a+}^{p}f(x)\right]=\mathfrak{D}_{a+}^{p+q}f(x)-\sum_{k=0}^{\mathbf{n}-1}\frac{(\mathfrak{D}_{a+}^{p-k-1}f)(a)}{\Gamma(-q-k)}(x-a)^{-q-k-1}, \quad (19)$$

where p, q > 0, $\mathbf{n} = [p] + 1$ and $\mathbf{m} = [q] + 1$.

Note that the equalities in Equation (18) and Equation (19) occurs if, and only if, the summands are nulls and this is guaranteed if

$$(\mathfrak{D}_{a+}^{\beta-k-1}f)(a)=0, \ (k=0,1,\ldots,\mathbf{n}-1), \quad (20)$$

$$(\mathfrak{D}_{a+}^{\alpha-k-1}f)(a)=0, \ (k=0,1,\ldots,\mathbf{m}-1). \quad (21)$$

The author [19] shows in section 2.3.7 of his book that if $f(x)$ have enough continuous derivatives, then the conditions in Equation (20) and Equation (21) are equivalent, respectively, to

$$f^{(k)}(a)=0, \ (k=0,1,\ldots,\mathbf{n}-1), \quad (22)$$

$$f^{(k)}(a)=0, \ (k=0,1,\ldots,\mathbf{m}-1). \quad (23)$$

Hence, if $f^{(k)}(a) = 0$, $(k = 0, 1, \ldots, \mathbf{r} - 1)$, where $\mathbf{r} = \max\{\mathbf{m},\mathbf{n}\}$, then

$$\mathfrak{D}_{a+}^{\alpha}\left[\mathfrak{D}_{a+}^{\beta}f(x)\right]=\mathfrak{D}_{a+}^{\alpha+\beta}f(x)=\mathfrak{D}_{a+}^{\beta}\left[\mathfrak{D}_{a+}^{\alpha}f(x)\right].$$

Finally, we also mention the following result

$$\mathfrak{D}_{a+}^{p}\left[\mathfrak{D}_{a+}^{-q}f(x)\right]=\mathfrak{D}_{a+}^{p-q}f(x), \ p,q\in\mathbb{R}^{+} \quad (24)$$

valid whenever $f(x)$ is at least continuous and when $p \geq q \geq 0$ that the derivative $\mathcal{D}_{a+}^{p-q}f(x)$ exists. Finalizing this section, we present a generalization of the Leibniz rule for integrodifferentiating the product of functions [14,16].

**Theorem 5 (Fractional Leibniz Rule):** Let $f$ and $g$ be analytical in $[a, b]$, then

$$\mathfrak{D}_{a+}^{\nu}[fg]=\sum_{k=0}^{\infty}\binom{\nu}{k}f^{(k)}\left(\mathfrak{D}_{a+}^{\nu-k}g\right), \ \nu\in\mathbb{R} \quad (25)$$

where $\binom{\nu}{k}$ are the generaliied binomial coeJlcients written in terms of the gamma function.

## Confluent Hypergeometric Equation

Inspired by a similar technique was developed [14] where we investigated how to obtain a solution of the Bessel's equation, in this section we verify the possibility of obtaining solutions of the confluent hypergeometric equation by means of fractional calculus methodology.

We start by considering the standard form of the confluent hypergeometric equation

$$x\frac{d^{2}u}{dx^{2}}+(c-x)\frac{du}{dx}-au=0, \quad (26)$$

defined for $x > 0$ with $a, c \in \mathbb{R}$. We now assume that for each $u = u(x)$ satisfying Equation (26), there exists an $f = f(x)$ differ integrable such that

$$u=\mathfrak{D}_{0+}^{-1-\alpha}f, \quad (27)$$

where α is an unknown parameter yet to be determined.

So rewriting Equation (26), it follows that

$$x\frac{d^{2}}{dx^{2}}\mathfrak{D}_{0+}^{-1-\alpha}f+(c-x)\frac{d}{dx}\mathfrak{D}_{0+}^{-1-\alpha}f-a\mathfrak{D}_{0+}^{-1-\alpha}f=0 \quad (28)$$

which by means of Eq.(24), assume the form

$$x\mathfrak{D}_{0+}^{1-\alpha}f+(c-x)\mathfrak{D}_{0+}^{-\alpha}f-a\mathfrak{D}_{0+}^{-1-\alpha}f=0, \quad (29)$$

and yet, by the action of $\mathfrak{D}_{0+}^{\alpha}$ applied to the left of each term in Eq.(29) can be rewritten as

$$\mathfrak{D}_{0+}^{\alpha}\left[x\mathfrak{D}_{0+}^{1-\alpha}f\right]+\mathfrak{D}_{0+}^{\alpha}\left[(c-x)\mathfrak{D}_{0+}^{-\alpha}f\right]-\mathfrak{D}_{0+}^{\alpha}\left[a\mathfrak{D}_{0+}^{-1-\alpha}f\right]=0. \quad (30)$$

Now, if we make use of the fractionalized version of the Leibniz rule as established by Theorem 5, it can be verified the following equalities

$$\mathfrak{D}_{0+}^{\alpha}\left[x\mathfrak{D}_{0+}^{1-\alpha}f\right]=x\frac{df}{dx}+\alpha f, \quad (31)$$

$$\mathfrak{D}_{0+}^{\alpha}\left[(c-x)\mathfrak{D}_{0+}^{-\alpha}f\right]=(c-x)f-\alpha\mathfrak{D}_{0+}^{-1}f, \quad (32)$$

whenever $f$ satisfies the following conditions

$$\left(\mathfrak{D}_{0+}^{-\alpha-k}f\right)(0)=0, \mathrm{k}=0,1,\ldots,[1-\alpha].$$

Hence, when Equation (31) and Equation (32) are substituted back in Equation (30) it returns

$$x\frac{df}{dx}+(\alpha+c-x)f+(-a-\alpha)\mathfrak{D}_{0+}^{-1}f=0. \quad (33)$$

Imposing that $\alpha = -a$, the last term of Eq.(33) vanishes and the equation reduces to the following equivalences

$$x\frac{df}{dx}+(\alpha+c-x)f=0 \Leftrightarrow \frac{1}{f}\frac{df}{dx}=1+\frac{(a-c)}{x} \quad (34)$$

which is a separable differential equation whose solution is given by

$$f(x)=Ke^{x}x^{a-c}, \quad (35)$$

with $K \in \mathbb{R}$. an arbitrary constant.

Finally, remembering the results of Equation (27) and again applying the fractional Leibniz rule we have that

$$u(x)=\mathfrak{D}_{0+}^{-1+a}\left[Ke^{x}x^{a-c}\right]$$

$$=K\sum_{n=0}^{\infty}\binom{-1+a}{n}\frac{d^{n}e^{x}}{dx^{n}}\mathfrak{D}_{0+}^{-1+a-n}\left[x^{a-c}\right]$$

$$=Ke^{x}\sum_{n=0}^{\infty}\binom{-1+a}{n}\frac{\Gamma(a-c+1)}{\Gamma(n+2-c)}x^{n-c+1}$$

$$=Kx^{1-c}e^{x}\sum_{n=0}^{\infty}\binom{-1+a}{n}\frac{\Gamma(a-c+1)}{\Gamma(n+2-c)}x^{n}. \quad (36)$$

The solution $u(x)$ just obtained does not resembles, at least a priori, to the canonical form ${}_1F_1(a;c;x)$ that we commonly encounter when solving the confluent hypergeometric equation by means of the Frobenius method, for example. However, it can be promptly modified by means of a sequence of algebraic manipulations as it follows. We begin by rewriting the binomial coefficients in term of the gamma function

$$\binom{-1+a}{n}=\frac{\Gamma(a)}{\Gamma(a-n)n!}. \quad (37)$$

Also, by the famous reflection formula

$$\Gamma(z)\Gamma(1-z)=\frac{\pi}{\sin\pi z}, \quad (38)$$

we have the identities

$$\Gamma(a-n)\Gamma(1-a+n) = \pi \sin[\pi(a-n)] = \pi(-1)^n \sin(\pi a). \quad (39)$$

Then, substituting Equation (37) and Equation (39) into the expression of Equation (36), we obtain

$$u(x) = K \frac{\Gamma(a)\Gamma(a-c+1)}{\pi \sin(\pi a)} x^{1-c} e^x \sum_{n=0}^{\infty} \frac{\Gamma(1-a+n)}{(-1)^n \Gamma(n+2-c)} \frac{x^n}{n!}. \quad (40)$$

Now if we make use of the Pochhammer symbols $(z)_n = \frac{\Gamma(z+n)}{\Gamma(z)}$ to write down

$$(1-a)_n = \frac{\Gamma(1-a+n)}{\Gamma(1-a)} \quad \text{and} \quad (2-c)_n = \frac{\Gamma(n+2-c)}{\Gamma(2-c)},$$

then

$$u(x) = K \frac{\Gamma(a)\Gamma(1-a)}{\pi \sin(\pi a)} \frac{\Gamma(a-c+1)}{\Gamma(2-c)} x^{1-c} e^x \sum_{n=0}^{\infty} \frac{(1-a)_n}{(2-c)_n} \frac{(-x)^n}{n!}, \quad (41)$$

nd yet, noticing that if n = 0 in Equation (39), then

$$\frac{\Gamma(a)\Gamma(1-a)}{\pi \sin(\pi a)} = 1,$$

and it follows that

$$u(x) = K \frac{\Gamma(a-c+1)}{\Gamma(2-c)} x^{1-c} e^x \sum_{n=0}^{\infty} \frac{(1-a)_n}{(2-c)_n} \frac{(-x)^n}{n!}. \quad (42)$$

With a trivial proper choice of the constant $K$ we may have the equality

$$K \frac{\Gamma(a-c+1)}{\Gamma(2-c)} = 1$$

such that our solution for the equation Equation (26) obtained by means of fractional calculus reduces to

$$\begin{aligned} u(x) &= x^{1-c} e^x \sum_{n=0}^{\infty} \frac{(1-a)_n}{(2-c)_n} \frac{(-x)^n}{n!} \\ &= x^{1-c} e^x {}_1F_1(1-a; 2-c; -x). \end{aligned} \quad (43)$$

Finally, using the following identity

$${}_1F_1(a;c;x) = e^x {}_1F_1(c-a;c;-x), \quad (44)$$

it is possible to recognize that

$${}_1F_1(a-c+1; 2-c; x) = e^x {}_1F_1(1-a; 2-c; -x)$$

so that solution u(x) is of the form

$$u(x) = x^{1-c} {}_1F_1(a-c+1; 2-c; x), \quad (45)$$

which we know to be a linear independent solution with ${}_1F_1(a;c;x)$ as long as $c \notin \mathbb{Z}$

Although this methodology does not explicit a second linear independent solution, once we have one of them the other can be found using, for example, the reduction of order method.

## Concluding Remarks

This paper discussed the solution of the confluent hypergeometric equation by means of a methodology associated with fractional calculus. Using the Riemann-Liouville formulation we have obtained one such solution in a very simple way. Comparing this proposed fractional methodology with the standard Frobenius one, we can argue in favor of the former by saying that although the classical Frobenius approach seems to be a more "natural approach", the computational effort of the fractional method seems to be much simpler. Also, the fractional method expose the possibility of highlighting (possibly new) non-trivial relations between the special functions of the mathematical physics, suggesting a fertile ground of study and research in this area. As previously mentioned in the end of the last section, for the second linearly independent solution one can, for example, use the reduction of order method. Although, at this moment, we do not envision a clear approach to get a second linearly independent solution using such fractional method, it is passive of further investigations. Nevertheless, the most important conclusion is that fractional calculus can be used to discuss an ordinary differential equation as an interesting alternative to obtain solutions and such mathematical tool should be encouraged to be investigated and used more frequently in theoretical and applied problems.